\documentclass[conference]{IEEEtran}
\IEEEoverridecommandlockouts
\usepackage{cite}
\usepackage{amsmath,amssymb,amsfonts}
\usepackage{algorithmic}
\usepackage{graphicx}
\usepackage{textcomp}
\usepackage{xcolor}
\usepackage[hyphens]{url}
\usepackage{hyperref}
\hypersetup{colorlinks=true,breaklinks=true}
\def\BibTeX{{\rm B\kern-.05em{\sc i\kern-.025em b}\kern-.08em
    T\kern-.1667em\lower.7ex\hbox{E}\kern-.125emX}}

\renewcommand{\baselinestretch}{0.89}
\begin{document}

\title{Intelligent Task Offloading: Advanced MEC Task Offloading and Resource Management in 5G Networks
}
\author{\IEEEauthorblockN{Alireza Ebrahimi}
\IEEEauthorblockA{\textit{Department of Electrical and Computer Engineering} \\
\textit{Clemson University}\\
Clemson, USA \\
alireze@clemson.edu}
\and
\IEEEauthorblockN{Fatemeh Afghah}
\IEEEauthorblockA{\textit{Department of Electrical and Computer Engineering} \\
\textit{Clemson University}\\
Clemson, USA \\
fafghah@clemson.edu}
\vspace*{-4cm}

}

\maketitle

\begin{abstract}
5G technology enhances industries with high-speed, reliable, low-latency communication, revolutionizing mobile broadband and supporting massive IoT connectivity. With the increasing complexity of applications on User Equipment (UE), offloading resource-intensive tasks to robust servers is essential for improving latency and speed. The 3GPP's Multi-access Edge Computing (MEC) framework addresses this challenge by processing tasks closer to the user, highlighting the need for an intelligent controller to optimize task offloading and resource allocation.
This paper introduces a novel methodology to efficiently allocate both communication and computational resources among individual UEs. Our approach integrates two critical 5G service imperatives: Ultra-Reliable Low Latency Communication (URLLC) and Massive Machine Type Communication (mMTC), embedding them into the decision-making framework. 
Central to this approach is the utilization of Proximal Policy Optimization, providing a robust and efficient solution to the challenges posed by the evolving landscape of 5G technology.
The proposed model is evaluated in a simulated 5G MEC environment. The model significantly reduces processing time by 4\% for URLLC users under strict latency constraints and decreases power consumption by 26\% for mMTC users, compared to existing baseline models based on the reported simulation results. These improvements showcases the model’s adaptability and superior performance in meeting diverse QoS requirements in 5G networks.
\end{abstract}

\vspace{-0.2cm}
\section{Introduction}
\vspace{-0.1cm}

The rapid advancement of Virtual Reality/Augmented Reality (VR/AR) and autonomous vehicles is transforming industries and everyday life. By 2028, the VR/AR market is projected to reach 3.67 billion users, generating about USD 58.3 billion, while the autonomous driving sector may achieve a market value of USD 13,632.4 billion with 54 million connected vehicles. These technologies require high data transfer rates, ultra-low latency, and robust security.

Despite improvements in computational capabilities of user equipment (UE), increasing application complexity poses challenges in meeting strict latency demands for optimal user experiences. Traditional cloud computing introduces latency and congestion, making it unsuitable for these technologies. Consequently, fog and edge computing have emerged as viable alternatives, with edge computing reducing latency by bringing computation closer to end devices \cite{fog}, \cite{survey}.

To effectively offload tasks to edge servers, a robust communication link is essential, offering broad coverage, low latency, high reliability, and security. Among wireless network technologies, 5G stands out as the most promising solution. Unlike WiFi, which is limited in range, 5G provides extensive coverage and high data rates \cite{5gvswifi}. The 3rd Generation Partnership Project (3GPP) classifies 5G services into three categories: Enhanced Mobile Broadband (eMBB), Ultra-Reliable Low Latency Communication (URLLC), and Massive Machine Type Communication (mMTC), each supporting a wide range of applications from remote surgery to massive IoT deployments \cite{5G}. The synergy between 5G and edge computing addresses the growing demand for low-latency, high-bandwidth applications, facilitating the proliferation of IoT devices, augmented reality, autonomous vehicles, and smart cities \cite{3gppwebsite,mec_5g_2,etsi1}.

The Multi-access Edge Computing (MEC) framework, introduced in 3GPP technical specifications, enables data processing and storage closer to end users, improving latency, reducing network congestion, and enhancing the overall user experience \cite{etsi2}. In urban environments, where URLLC and mMTC users have varying offloading requests, efficient resource allocation becomes crucial, necessitating a balance between limited communication and computation resources. The Open Radio Access Network (O-RAN), a non-proprietary version of the Radio Access Network (RAN), integrates MEC platforms within the RAN infrastructure, providing the computational resources needed for task offloading. O-RAN's modular design introduces intelligent controllers, such as the Near-Real-Time Intelligent Controller (Near-RT RIC) and the Non-Real-Time Intelligent Controller (Non-RT-RIC), which optimize resource allocation and prioritize critical tasks within the xApp framework of the O-RAN architecture \cite{arnaz}.

However, existing research has not fully addressed the consideration of diverse network slices, including URLLC and mMTC users, particularly in the context of task offloading and resource allocation. While some studies focus on the allocation of only communication resources among UEs \cite{fatemeh}, others that addressed both communication and computation allocation, often overlook the distinct requirements of different network slices \cite{joint} which is very important due to intrinsic difference between their tasks. This research gap warrants further exploration.

In this study, we develop intelligent models for task  offloading as well as allocation of communication and computational resources using  Deep-Q Learning and Proximal Policy Optimization (PPO) techniques. Our approach significantly outperforms established baselines, demonstrating its effectiveness and efficiency.

The subsequent sections of this paper are structured in the following manner: Section \ref{rel} delves into a review of the existing literature and previous works that are pertinent to our study. Following this, Section \ref{sys} introduces the system model that we have adopted for our study. In Section \ref{AA}, we present the problem formulation and the optimization. Section \ref{sim} is dedicated to the presentation and discussion of the simulation results. Finally, Section \ref{con} concludes the paper.

\vspace{-0.2cm}
\section{Related Works}
\vspace{-0.15cm}
\label{rel}


In this section, we will first review advancements in task offloading on MEC. Next, we will explore scholarly works focused on communication resource allocation in 5G networks. Finally, we will examine studies that integrate both communication and computation resource allocation for task offloading in 5G. This systematic approach will offer a comprehensive understanding of the current research landscape in these areas.



\subsection{Task Offloading in Edge}
In the study presented in \cite{DAG}, the researchers introduce a heuristic algorithm, referred to as the Table Based Task Offloading Algorithm (TBTOA). This algorithm addresses the issue of task offloading in Mobile Edge Computing (MEC), taking into account both dependency and service caching constraints.
The authors of \cite{vec} delve into the problem of dependency-aware task offloading and service caching within the context of vehicular edge computing. This is a significant application of MEC within the realm of intelligent transportation systems. A similar investigation into MEC task offloading for vehicles is conducted in \cite{vec2}.
The paper \cite{tradeoff} explores the balance between delay and energy consumption in the task offloading process within a multi-user MEC scenario. The authors provide insights on the decision-making process regarding whether a task should be offloaded for edge execution.
In \cite{edge1} the  authors use edge servers as a processing platform to assist UAVs to offload their signal processing tasks. 

\subsection{Resource Block Allocation in 5G}
The study denoted as \cite{fatemeh} showcases the use of Long Short-Term Memory (LSTM) networks by researchers for predicting network traffic. This was aimed at network slicing for various 5G services, each with distinct Quality of Service (QoS) levels, with deep Q-learning employed for decision-making.
Attention-based deep reinforcement learning (ADRL) technique was proposed by the authors in \cite{fatemeh_2} for the optimal control of dynamic 5G networks.
Federated learning and team learning were utilized for resource allocation in the research conducted by \cite{team}.

\subsection{Edge Computing in 5G}

The exploration of URLLC and eMBB slices was the focus of \cite{vnf}. The authors aimed to decompose base station functions into Virtual Network Functions (VNFs) and utilized federated deep Q-learning for distributed learning, aiming to optimize resource utilization at edge sites while minimizing network reconfiguration errors.


The publication \cite{2018} addresses concerns regarding execution delay and energy consumption when deciding on offloading requests. They take into account factors such as task data volume, required CPU cycles for data processing, and maximum tolerable task delay, utilizing deep Q-learning to determine which tasks should be offloaded.


The authors of \cite{ee} proposed a time window for wireless power transfer and task offloading from IoT devices to Multi-access Edge Computing (MEC) servers. They introduced the Deep Reinforcement Learning-based Online Offloading (DROO) framework, employing a deep neural network for scalable decision-making based on past experiences. An adaptive procedure is incorporated to adjust algorithm parameters dynamically.


Intelligent Ultradense Edge Computing (I-UDEC), introduced in \cite{when}, integrates MEC servers and device-to-device communication. The aim was to minimize overall task processing time. The framework determines task processing locations, resource allocation strategies, and identifies services suitable for caching on edge servers to reduce computational time.


A novel paradigm called Sensing-Assisted Wireless Edge Computing (SAWEC) was introduced in \cite{sawec} to improve the performance of mobile virtual reality (VR) systems by leveraging knowledge about the physical environment and transmitting only the relevant data for service delivery to the edge server.


The authors of \cite{UAV} presented a novel A2-UAV framework for optimizing task execution in multi-hop UAV networks, considering factors like deep neural network accuracy, image compression, target positions, and UAV energy position. This framework significantly outperforms existing solutions.

The authors of \cite{joint}, the authors explore joint slicing of both communication and computation resources in a Radio Access Network (RAN), employing a two-tier slicing approach. The first tier, computation slicing, determines task execution locations and allocates computation resources using deep Q-learning. The second tier involves communication resource slicing, allocating communication resource blocks also through the application of deep Q-learning.


\section{System Model}
\label{sys}
\begin{figure}
    \centering
    \includegraphics[width=1\linewidth]{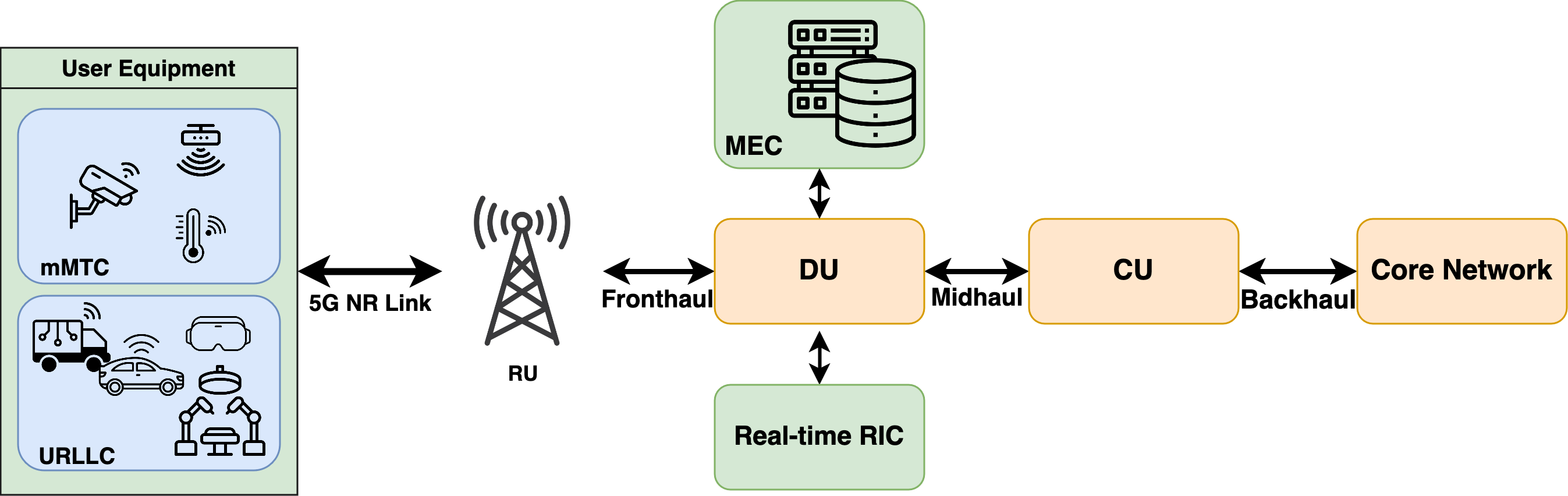}
    \caption{System Model}
    \label{sysmodel}
\end{figure}

In this paper, we examine an Open Radio Access Network (O-RAN) architecture that incorporates a single Radio Unit (RU). This RU communicates with users via radio communication. The signals received are subsequently transmitted to the Distributed Unit (DU) via the network’s fronthaul. The DU is outfitted with a Multi-access Edge Computing (MEC) server, which is responsible for processing offloaded tasks. A real-time RAN Intelligent Controller (RIC) is also housed within the DU, where it operates xApps. Communication between the DU and the Centralized Unit (CU) is facilitated through the midhaul. Lastly, the CU is linked to the core network via the network’s backhaul. This comprehensive setup forms the basis of our O-RAN network structure under consideration and is shown in figure \ref{sysmodel}\cite{3gpp-arch}.

The set of services \(S\) in the network contains two slice types of URLLC and eMBB (\(S = \{1, 2\}\)).  
The total number of users connected to the RU is \(N\) with \(N_s\) UEs connected to the RU with \(s \in S\).
Each user equipment $j$ of slice \(s\) connected to the RU, denoted by \(u_{j, s}\) randomly generates a task \(T_j\). Each task could be defined as a tuple (\(b_j\), \(c_j\), \(\tau_j\)) with \(b_j\) be the number of bytes that should be offloaded, \(c_j\) be the number of CPU cycles needed to process that task, and \(\tau_j\) denotes the acceptable latency for processing the task. The UEs are scattered in the coverage area of the RU, where  \(d_j\) denotes the distance between UE $j$ and the RU. 

The RU has limited communication and computational resources that can be allocated to UEs. The total number of communication resource blocks (RBs) is denoted as \(R_{comm}\). Similarly, computational resources of the MEC server, \(R_{comp}\), can be allocated to UEs for task processing, with multiple resources reducing processing time. Table \ref{nota} shows all the notations with their description used in this paper.

\begin{table}[]
\begin{center}
\caption{Notation Description}
\label{nota}
\begin{tabular}{|c|l|}
\hline
\textbf{Parameter}                       & \multicolumn{1}{c|}{\textbf{Description}}               \\ \hline
$N$                                        & Total number of users                                   \\ \hline
$N_s$                                     & Number of users in slice s                            \\ \hline
$S$                                        & Set of services offered                                 \\ \hline
$s$                                        & UE slice                                                \\ \hline
$j$                                        & UE                                                      \\ \hline
$u_{j, s}$                               & User j from slice s                                     \\ \hline
$T_j$                                     & Task generated by UE j                                  \\ \hline
$b_j$                                     & Number of bytes from task $T_j$                          \\ \hline
$c_j$                                     & Number of CPU cycles from task $T_j$                     \\ \hline
$\tau$                                     & Acceptable latency for task $T_j$                        \\ \hline
$d_j$                                     & Distance between UE j and RU                            \\ \hline
$R_{comm}$                                 & Total number of communication RBs                       \\ \hline
$R_{comp}$                                 & Total number of computational resources                 \\ \hline
$C_j$                                    & Channel capacity between UE j and RU                    \\ \hline
$K_j^{comm}$                              & Number of communication RB allocated to UE j            \\ \hline
$K_j^{comp}$                               & Number of computation resources allocated to UE j       \\ \hline
$B$                                        & Bandwidth of each communication RB                      \\ \hline
$P_j^{trans}$                              & Transmission power of UE j                              \\ \hline
$P_{local, j}^{process}$ & Processing power when the task is processed locally     \\ \hline
$P_j^{idle}$           & Idle power when the task is processed on MEC            \\ \hline
$\eta$                                      & Path loss exponent                                      \\ \hline
$|h_j|^2$               & Time-varying Rayleigh fading channel gain               \\ \hline
$\sigma^2_n$                 & Variance of the noise                                   \\ \hline
$t_j^{trans}$      & Transmission delay of task $T_j$                         \\ \hline
$f_s$                                     & CPU frequency of UE in slice s                          \\ \hline
$f_{rb}$                                    & CPU frequency of each computational resources           \\ \hline
$t_{local, j}^{process}$                     & Processing time of task $T_j$ locally                    \\ \hline
$t_{MEC, j}^{process}$                      & Processing time of task $T_j$ on MEC                     \\ \hline
$E_{local, j}^{process}$                     & Energy consumption for processing the task $T_j$ locally \\ \hline
$E_{MEC, j}^{process}$                      & Energy consumption for processing the task $T_j$ on MEC  \\ \hline
$\gamma$                                    & Discount factor                                         \\ \hline
$\alpha, \beta, \delta$                                    & Weights associated with each term in the reward function  \\ \hline
$t_{exe}$                                  & Time of executing task $T_j$                             \\ \hline
$E_{exe}$                                   & Energy consumption of task $T_j$                         \\ \hline
\end{tabular}
\end{center}
\vspace{-0.9cm}
\end{table}


\subsection{Wireless Communication}
 The channel capacity for the orthogonal frequency-division multiple access (OFDMA) channel between the UE and base station is as follows:
 \vspace{-0.45cm}
\begin{equation}
C_j = K_j^{comm} \times B \times log_2(1 + \frac{P_j^{trans}\: d_j^{-\eta}\:|h_j|^2}{\sigma^2_n})
\vspace{-0.2cm}
\end{equation}
where \(C_j\) is the channel capacity between user \(j\) and the RU, 
\(K_j^{comm}\) is the number of communication RBs from the RU that has been assigned to user \(j\), \(B\) is the RB bandwidth, \(P_{j, trans}\) is the transmission power of user \(j\), \(\eta\) is the path loss exponent, \(|h_j|^2\) is the time-varying Rayleigh fading channel gain and \(\sigma^2\) represents the variance of the noise.

The transmission delay for offloading the task \(T_j\) from UE to the MEC server can be expressed as:
\begin{equation}
    t_{j}^{trans} = \frac{b_j}{C_j}
\end{equation}

\subsection{Computational Resources}
For each task there is a specific number of CPU cycles needed to process it. The duration of this processing depends on the frequency of the CPU. Each UE based on the slice that it is in would have the frequency of \(f_s\). Also for the MEC server the frequency of each RB is \(f_{rb}\). Assuming that the task \(T_j\) will be offloaded to the MEC server and \(K_j^{comp}\) number of computational resources will be assigned for processing this task, the equivalent frequency for processing this task can be expressed as follows:
\vspace{-0.35cm}
\begin{equation}
    f_m = K_j^{comp} \times f_{rb}
    \vspace{-0.2cm}
\end{equation}

Therefore, the processing time based on the processing location can be expressed as follows:
\begin{subequations}
\begin{align}
    t^{process}_{local, j} = \frac{c_j}{f_s} \\ 
    t_{MEC, j}^{process} = \frac{c_j}{f_m}
\end{align}
\end{subequations}

\subsection{Energy Consumption}
Energy consumption is a critical consideration for IoT devices due to their reliance on continuous operation and often limited power sources. Efficient energy usage not only prolongs device lifespan but also minimizes maintenance requirements and operational costs. Moreover, optimizing energy consumption enhances the scalability and reliability of IoT networks, ensuring seamless connectivity and functionality across diverse environments.

Energy calculation for each UE when the task is processed locally and when it is perocessed on the MEC server can be expressed respectively as follows:
\begin{subequations}
\begin{align}
    &E^{process}_{local, j} = P^{process}_{local, j} \times  t^{process}_{local, j}\\ 
    &E_{MEC, j}^{process} = P^{trans}_j \times  t^{trans}_j + P^{idle}_j \times  t_{MEC, j}^{process}
\end{align}
\end{subequations}
where \(P^{process}_{local}\) is the device power consumption during processing the task locally, \(P^{trans}\) is the power consumption during sending the task to the server and \(P^{idle}\) is the device power consumption while waiting for the task to be processed on the server. Since \(P^{idle}\) is negligible, it is considered to be zero.
\vspace{-0.3cm}
\section{Proposed Method and Optimization}\label{AA}
\vspace{-0.15cm}
In the proposed MEC-equipped system, the volume of offloading task requests from users frequently exceeds the available resources, necessitating an intelligent controller to manage task offloading and resource distribution effectively. 
The main objective of this optimization is to minimize the total execution time while considering the unique requirements of users belonging to different network slices. For URLLC users, the critical factor is ensuring that processing times do not exceed the predefined latency limits, as any violation would breach the Service Level Agreement (SLA). 
For mMTC, the focus shifts to minimizing the energy consumption since these tasks are typically handled by battery-powered IoT devices. The intelligent controller must make real-time decisions on which tasks to offload based on the current system load and available resources. This involves prioritizing tasks by slice type and urgency, dynamically allocating computational resources, and adjusting strategies as network conditions change.

In a Markov Decision Process (MDP), the state space represents all possible system conditions, defined by current load, available resources, and the task requesting offloading. The action space includes decisions like task offloading, task prioritization by slice type, and resource allocation. Transition probabilities indicate the likelihood of moving between states based on actions, adhering to the Markov property. The reward function quantifies the desirability of state-action pairs, encouraging decisions that minimize execution time.

Reinforcement Learning (RL) helps the controller adapt to changes in a dynamic environment, handling delayed rewards. It manages the trade-off between exploration (testing new strategies) and exploitation (using known methods). The goal is to find an optimal policy that minimizes execution time while meeting network slice requirements, making RL ideal for this scenario.
To address this problem, we propose a Proximal Policy Optimization (PPO)-based task offloading and joint resource allocation strategy. PPO is selected due to its ability to handle large and continuous state spaces, which is essential for efficiently managing dynamic resource allocation in a 5G multi-access edge computing environment. The agent learns to assign resources among users with task offloading requests during the training process.
The reward function that the agent is trying to maximize is as follows:
\vspace{-0.2cm}
\begin{equation}
    R = \sum_{t=1}^{\infty} \gamma^{t} \sum_{j=1}^{N} w_s \:r_{j, t}(k^{comp}, k^{comm})
    \vspace{-0.4cm}
\end{equation}
s.t.:
\begin{subequations}
\begin{align}
    \sum_{j=1}^{N} K_j^{comm} \leq R_{comm} \label{c1}\\ 
    \sum_{j=1}^{N} K_j^{comp} \leq R_{comp} \label{c2}
\end{align}
\end{subequations}
where \(\gamma\) is the discount factor, \(w_s\) is the weight associated with each slice, \(r_{j,t}(k^{comp}, k^{comm})\) is the reward from each user after assigning \(k^{comm}\) communication resource and \(k^{comp}\) computation resource. Constraints \ref{c1} and \ref{c2} make sure that the number of assigned communication and computational resources to the users do not exceed the available resources respectively. 

The observation for the task \(T_j\) from user \(u_{j,s}\) is defined as \(obs=(s, b_j, c_j, \tau_j, C_j, R_{comm}^{rem}, R_{comp}^{rem}, f_s, f_{rb})\), where \(R_{comm}^{rem}\) and \(R_{comp}^{rem}\) are the remaining communication  and computational resources when task \(T_j\) arrives. The action is also defines as \(action=(K_j^{comm}, K_j^{comp})\).

In case of deciding that a task should be processed locally, the agent would not assign any resource to that task therefore \(K^{comp}=K^{comm}=0\). For that reason binary variable \(x_j\) is defined to show whether that task should be processed locally (\(x_j = 0\)) or offloaded to the MEC server (\(x_j = 1\)).


\subsection{URLLC Reward Function}
As mentioned previously, there are two important criteria for URLLC users while addressing the offloading decision. The first one is to minimize the execution time and the second one enforces not violating the SLA by the latency of processing the task.
The following is the reward function for URLLC users:
\vspace{-0.4cm}
\begin{equation}
\begin{split}
    r_{URLLC} = \frac{2}{1+e^{-\delta \times r_{u}(t_{exe})}} - 1
\end{split}
\end{equation}
\vspace{-0.5cm}
where,
\begin{equation*}
\begin{split}
    r_{u}(t_{exe}) =  &\alpha \times (\frac{t^{process}_{local} - t_{exe}}{t^{process}_{local}}) + \beta \times (\frac{\tau - t_{exe}}{\tau}),
\end{split}
\end{equation*}
\begin{equation*}
\label{t_exe}
    t_{exe} = \left\{\begin{matrix}
t^{process}_{local} & x_i = 0\\ 
t^{trans} + t_{MEC}^{process} & x_i = 1
\end{matrix}\right.
\end{equation*}
where \(t_{exe}\) is the latency of processing the task. If there are no resources assigned to this task, then the task is processed locally, otherwise the task would be offloaded to MEC server.
\subsection{mMTC Reward Function}
Since most of the devices using this service, are battery powered IoT devices, then in addition to decreasing the latency of processing the data, decreasing the power consumption is also of importance. Therefore, the reward function for this service is as follows:
\vspace{-0.4cm}
\begin{equation}
\begin{split}
    r_{mMTC} = \frac{2}{1+e^{-\delta \times r_{m}(t_{exe}, E_{exe})}} - 1
\end{split}
\end{equation}
\vspace{-0.4cm}
where, 
\begin{equation*}
\begin{split}
    r_{m}(t_{exe},E_{exe}) = &\alpha \times (\frac{t^{process}_{local} - t_{exe}}{t^{process}_{local}}) \\&+ \beta \times  \frac{E_{local}^{process} - E_{exe}}{E_{local}^{process}},
\end{split}
\end{equation*}
\begin{equation*}
\label{E_exe}
    E_{exe} = \left\{\begin{matrix}
E_{local}^{process} & x_i = 0\\ 
E_{MEC}^{process} & x_i = 1
\end{matrix}\right.
\end{equation*}
where \(E_{exe}\) is the energy consumption of processing the task.

\section{Simulation Results}
\label{sim}


The simulation environment is comprised of a centrally positioned base station, covering a rectangular area of \(2000 \times 3000\) meters. This base station is equipped with a MEC, a high-performance server featuring a 4-core CPU operating at a frequency of 2 GHz. The server has the capacity to allocate up to 40 units of computational resources to the users. Concurrently, the radio unit, is equipped to supply 80 blocks of communication resources. 

The number of UEs for each network slice that generate tasks at each time step is modeled as a random variable following a Gaussian distribution. The environment represents an urban setting with a higher density of IoT devices than UEs. Specifically, the number of URLLC UEs have a mean (\(\mu\)) of 10 and a variance (\(\sigma^2\)) of 2, while the mMTC UEs have a mean of 30 and a variance of 5. These UE are uniformly distributed within the coverage area.
The communication channel consists of 80 resource blocks, each with a bandwidth of 4 MHz, which can be dynamically allocated to users based on their requirements. The channel conditions are modeled using a Rayleigh fading model Detailed parameters for the environment are provided in Table \ref{env_param}.

Each UE generates a task that requires \(b\) bytes and \(c\) CPU cycles, following a uniform distribution within the bounds specified in Table \ref{ue_param}. The CPU frequency of each UE is contingent upon whether they belong to the URLLC or mMTC slice. Similarly, the power consumption of each UE is dependent on the slice to which they belong. The parameters used for the UEs are outlined in Table \ref{ue_param}.

The PPO agent employs a neural network architecture consisting of four hidden layers with 128, 256, 128, and 64 neurons, respectively. The model was trained for 15,000 episodes to optimize policy performance through iterative updates.
The specific hyperparameters employed during the training phase are detailed in Table \ref{hyper_params}.


Simulation results averaged over 5,000 experiments, illustrating the variations in performance metrics with varying numbers of UEs, and communication channels for RL agents compared to sequential and fair assignments. 

\begin{figure*}
\vspace{-0.1cm}
    \centering
    \includegraphics[width=1\linewidth]{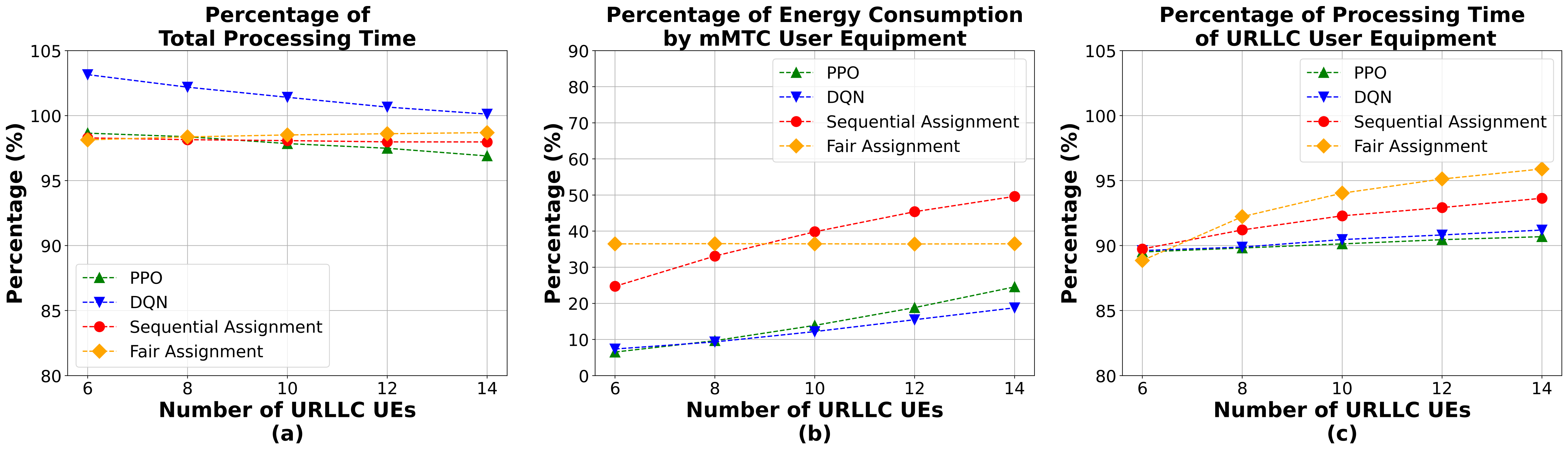}
    \vspace*{-7mm}
    \caption{Simulation results illustrating the variation in performance metrics with varying numbers of URLLC UEs, alongside 30 mMTC UEs, for RL agents compared to sequential and fair assignments. (a) Percentage of total processing time as the number of URLLC UEs increases. (b) Percentage of energy consumption by mMTC UEs as the number of URLLC UEs increases. (c) Percentage of total processing time for URLLC UEs.}
\vspace{-0.4cm}
    \label{aggaa_1_nonforbiden}
\end{figure*}

Figure \ref{aggaa_1_nonforbiden} demonstrates the variations in three critical performance indicators as the number of URLLC UEs increases, while the number of mMTC users remains constant at 30. 
The proposed PPO method was compared with another RL approach, Deep Q-Learning (DQN), as well as two additional baseline methods. The DQN agent utilized the same neural network architecture as the PPO agent, with identical training parameters and the same number of epochs (15,000) for training.
The first baseline, referred to as the sequential baseline, assigns resources to each incoming request, given that resources are available. If the request is from a URLLC UE, the allocated resource must meet the latency requirement. On the other hand, if the request is from an mMTC UE, the allocated resource should result in a decrease in energy consumption compared to local processing. However, this method does not guarantee fairness between the two slices. If all URLLC requests are received before mMTC offloading requests, all resources would be allocated to URLLC users, thereby contravening the principle of fairness.

An alternative approach involves predetermining the proportion of resources to allocate to each slice before allocation. This is the strategy used in the second baseline, where available resources are evenly distributed between each slice. This method ensures a more equitable distribution of resources, fostering fairness between the slices.

Subfigure \ref{aggaa_1_nonforbiden}a illustrates the percentage of total processing time required to handle tasks using different approaches, relative to processing them entirely locally, as the number of URLLC users varies. For instance, the fair assignment approach achieves 98\% in total processing time, indicating a 2\% reduction in processing time when tasks are offloaded and resources are assigned according to this approach, compared to the scenario where all tasks are processed locally without offloading.

Subfigure \ref{aggaa_1_nonforbiden}b shows the energy consumption percentage for mMTC users as the number of URLLC users changes.

Subfigure \ref{aggaa_1_nonforbiden}c demonstrates the proportion of processing time for URLLC UE tasks as the number of URLLC users varies.

\begin{figure*}

    \centering
    \includegraphics[width=1\linewidth]{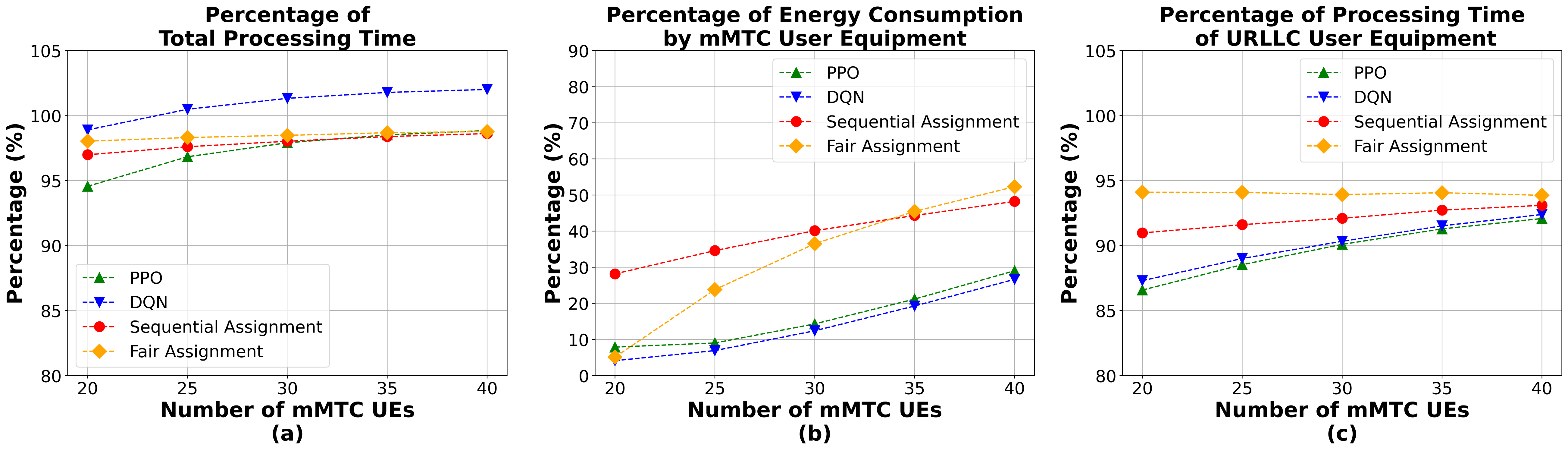}
    \vspace*{-7mm}
    \caption{Simulation results depicting the variation in performance metrics with varying numbers of mMTC UEs, alongside 10 URLLC UEs compared to sequential assignment and fair assignment. (a) Percentage of total processing time as the number of mMTC UEs increases. (b) Percentage of energy consumption by mMTC UEs. (c) Percentage of total processing time for URLLC UEs as the number of mMTC UEs increases}
    \label{aggaa_3_nonforbiden}
    \vspace*{-7mm}
\end{figure*}

Figure \ref{aggaa_3_nonforbiden} illustrates the fluctuation in the aforementioned performance indicators as the number of mMTC UEs escalates, while the count of URLLC users remains static at 10. This is compared against the baselines discussed earlier.

As observed, both reinforcement learning agents outperform the baseline methods in reducing power consumption and processing time for URLLC UEs. However, the DQN agent exhibits an increase in total processing time, primarily due to the additional delay introduced when offloading data to the MEC server, which leads to longer processing times compared to local processing. This effect is particularly noticeable for mMTC UEs, where one of the key objectives is to minimize power consumption. In contrast, this increase in processing time is not observed with the PPO agent, which also outperforms the DQN agent. The PPO agent not only surpasses both baselines in reducing total processing time but also achieves this while only marginally increasing power usage compared to the DQN agent.
Our methodology demonstrate optimal performance across all evaluated metrics. The PPO yield comparable results in reducing overall processing time (Figures \ref{aggaa_1_nonforbiden}a and \ref{aggaa_3_nonforbiden}a), power consumption (Figures \ref{aggaa_1_nonforbiden}b and \ref{aggaa_3_nonforbiden}b), and URLCC processing time (Figures \ref{aggaa_1_nonforbiden}c and \ref{aggaa_3_nonforbiden}c). The fair assignment method shows consistent performance irrespective of variations in the number of URLLC UEs concerning the energy consumption of mMTC UEs (Figure \ref{aggaa_1_nonforbiden}b), and it remains stable with changes in the number of mMTC UEs regarding the processing time of URLLC UEs. Conversely, the sequential assignment method changes with respect to the number of UEs.
Given the finite nature of resource allocation, a lower number of requests results in a higher acceptance rate. Consequently, the overall processing time, power usage by mMTC users, and latency for URLLC users are significantly reduced compared to scenarios where offloading is not an option.
However, as the user count—and therefore the request count—increases, the allocation challenge intensifies for the agent. This leads to a substantial decrease in the number of accepted task offloading requests and therefore decrease in overall performance. 
The simulation results depicted in Figures \ref{aggaa_1_nonforbiden} and \ref{aggaa_3_nonforbiden} present a comprehensive analysis of the system’s behavior for varying number of UEs.

\begin{table}[]
\begin{center}
\caption{Experimental Settings for Environment}
\vspace{-0.3cm}
\label{env_param}
\begin{tabular}{|c|c|}
\hline
\textbf{Parameters}                & \textbf{Values}   \\ \hline
MEC CPU Frequency                 & 4 $\times$ 2 GHz \\ \hline
Resource Block Bandwidth          & 4 MHz            \\ \hline
\(\eta\)                               & 2.8              \\ \hline
Number of Communication Resources & 80              \\ \hline
Number of Computational Resources & 40              \\ \hline
\end{tabular}
\end{center}
\end{table}
\vspace{-0.3cm}
\begin{table}[]
\vspace{-0.7cm}
\begin{center}
\caption{Experimental Settings for User Equipment}
\vspace{-0.3cm}
\label{ue_param}
\begin{tabular}{|c|cc|}
\hline
\textbf{Parameters} & \multicolumn{1}{c|}{\textbf{URLLC}}                          & \textbf{mMTC}         \\ \hline
Number of UE        & \multicolumn{1}{c|}{\(N(10, 2)\)}                            & \(N(30, 5)\)        \\ \hline
CPU Frequency       & \multicolumn{1}{c|}{600 MHz} & 200 MHz               \\ \hline
Transmission Power  & \multicolumn{1}{c|}{200 mW}                                  & 200 mW                \\ \hline
Processing Power    & \multicolumn{1}{c|}{-}                                       & 400 mW                \\ \hline
Task Bytes          & \multicolumn{1}{c|}{{[}2, 5{]} MB}                   & {[}2, 5{]} MB \\ \hline
Task CPU Cycles & \multicolumn{1}{c|}{{[}180, 660{]} \(\times\) \(10^6\)}                & {[}60, 220{]} \(\times\) \(10^6\)               \\ \hline
\(\tau\)            & \multicolumn{1}{c|}{700 ms}                                  & -                     \\ \hline
Weights         & \multicolumn{2}{c|}{\begin{tabular}[c]{@{}c@{}}$\alpha = 0.5, \beta = 0.5, \delta = 3$\end{tabular}} \\ \hline
\end{tabular}
\end{center}
\vspace{-0.9cm}
\end{table}
\begin{table}[]
\vspace{-0.3cm}
\begin{center}
\caption{Experimental Hyper-Parameters}
\vspace{-0.3cm}
\label{hyper_params}
\begin{tabular}{|c|c|}
\hline
\textbf{Parameters} & \textbf{Values} \\ \hline
Learning Rate       & 0.0001          \\ \hline
Batch Size          & 32              \\ \hline
Discount Factor     & 0.99            \\ \hline
Hidden Size         & 128, 256, 128, 64              \\ \hline
Hidden Layers       & 4              \\ \hline
\end{tabular}
\end{center}
\vspace{-0.2cm}
\end{table}
\section{Conclusion}
\label{con}
This study presents an end-to-end solution for ultra low latency task offloading at the ORAN-enabled 5G edge, addressing diverse emerging services including autonomous driving and low-power IoT. The reinforcement learning-based solution incorporate the best offloading strategy along with the optimized allocation of the required communication and computing resources to enhance overall performance for each network slice URLLC, mMTC). The decision-making process is powered by Deep Q-Learning and Proximal Policy Optimization.
In this methodology, an agent is trained on a diverse set of user equipment, each with varying task bytes, CPU cycles, and requirements that are contingent on the network slice they are associated with. This dynamic and adaptable training process equips the agent with the ability to handle a wide range of scenarios and conditions.
The simulation results underscore the effectiveness of  Proximal Policy Optimization as a decision-making tool in this context. It provides compelling evidence that our approach can successfully navigate the complexities of resource allocation in a multi-slice network environment, thereby validating the potential of our proposed methodology.
\vspace{-0.3cm}
\def\baselinestretch{0.82}
\bibliographystyle{IEEEtran}
\bibliography{IEEEabrv,ref}
\end{document}